\documentclass[9pt]{article}

\usepackage[utf8]{inputenc}
\usepackage[english]{babel}
\usepackage{graphics}
\usepackage{graphicx}
\usepackage{lipsum}
\usepackage[squaren, Gray, cdot]{SIunits}
\usepackage{epstopdf}
\usepackage{color}
\usepackage{transparent}
\usepackage{rotating}
\usepackage{bbm}
\usepackage{amsmath,amsfonts,amsthm,amssymb}
\usepackage{xcolor}
\usepackage{pgfplots}
\usepackage{eurosym}
\usepackage{wrapfig}
\usepackage{enumitem}
\usepackage[colorlinks=true,urlcolor=blue,linkcolor=blue,citecolor=blue]{hyperref}
\usepackage{multicol}
\usepackage{subfig}
\usepackage[countmax]{subfloat}

\addto\captionsenglish{}
\addto\captionsenglish{}

\usepackage{xspace} 
\renewcommand\Re{\operatorname{Re}}

\usepackage[top=2cm, bottom=2cm, left=2cm, right=2cm]{geometry}

\usepackage{fancyhdr}
\title{Decoupling of hyperfine structure of Cs $D_1$ line in strong magnetic field studied by selective reflection from a nanocell}

\author{Armen Sargsyan$^1$, Emmanuel Klinger$^{1,2}$, Grant Hakhumyan$^{1}$, Ara Tonoyan$^{1,2}$,  Aram Papoyan$^{1}$,\\
 Claude Leroy$^{2}$, David Sarkisyan$^{1}$ \\
\small$^1$ Institute for Physical Research, NAS of Armenia, Ashtarak-2, 0203 Armenia\\
\small$^2$ Laboratoire Interdisciplinaire Carnot de Bourgogne, UMR CNRS 6303 - Universit{\'e} Bourgogne, Franche-Comt{\'e},\\
\small BP 47870, 21078 Dijon Cedex, France \\
}

\begin{document}

    \maketitle
    \begin{center}
    \rule{14.5cm}{0.5pt}
 \section*{Abstract}
 \end{center}

\paragraph{} Decoupling of total electronic and nuclear spin moments of Cs atoms in external magnetic field for the case of atomic $D_1$ line, leading to onset of the hyperfine Paschen-Back regime has been studied theoretically and experimentally. Selective reflection of laser radiation from an interface of dielectric window and atomic vapor confined in a nanocell with 300 nm gap thickness was implemented for the experimental studies. The real time derivative of selective reflection signal with a frequency position coinciding with atomic transitions was used in measurements, providing $\sim$ 40 MHz spectral resolution and linearity of signal response in respect to transition probability. Behavior of 28 individual Zeeman transitions in a wide range of longitudinal magnetic field (0 -- 6 kG) has been tracked under excitation of Cs vapor by a low-intensity $\sigma^+$- polarized cw laser radiation. For $B\ge 6~$kG, only 8 transitions with nearly equal probabilities and the same frequency slope remained in the spectrum, which is a manifestation of the hyperfine Paschen-Back regime. The obtained experimental results are consistent with numerical modeling. Due to small divergence of selective reflection signal, as well as sub-wavelength thickness and sub-Doppler spectral linewidth inherent to nanocell, the employed technique can be used for distant remote sensing of magnetic field with high spatial and $B$-field resolution.
\vspace{0.5cm}
\begin{center}
\rule{14.5cm}{0.5pt}
\end{center}
\vspace{1cm}

\vspace{1cm}
\section{Introduction}

Optical nanometric thin cell (nanocell) containing atomic vapor of alkali metal (Rb, Cs, K, Na) is proven to be efficient and convenient spectroscopic tool for magneto-optical studies of optical atomic transitions between the hyperfine levels in strong external magnetic fields. Two interconnected effects develop with the increase of $B$-field: strong deviation of Zeeman splitting of hyperfine transitions from linear dependence, and significant change in probability of individual Zeeman transitions \cite{Tremblay,Aleksandrov,Auzinsh}. The efficiency of nanocell technique for quantitative spectroscopy of the Rb atomic transitions in strong magnetic field (up to 7 kG) resulting in onset of the hyperfine Paschen-Back regime was demonstrated in \cite{Sargsyan1,Sargsyan2,Sargsyan3}. The presence of specific "guiding" transitions foretelling characteristics of all other transitions between magnetic sublevels of alkali atoms $D_1$ line excited by $\pi$-polarized radiation in a strong transversal magnetic field was recently shown in \cite{Sargsyan4}. Micro- and nanocells were used in \cite{Sargsyan5} to study the onset of hyperfine Paschen-Back regime for Cs $D_2$ line in magnetic fields up to 9 kG. Realization of a complete hyperfine Paschen-Back regime in relatively weak magnetic field ($\sim$ 1 kG) was obtained in potassium nanocell on $D_1$ line of $^{39}$K in \cite{Sargsyan6}.

\begin{figure}[tb]
	\centering
	\begin{center}
		\includegraphics[width=345pt]{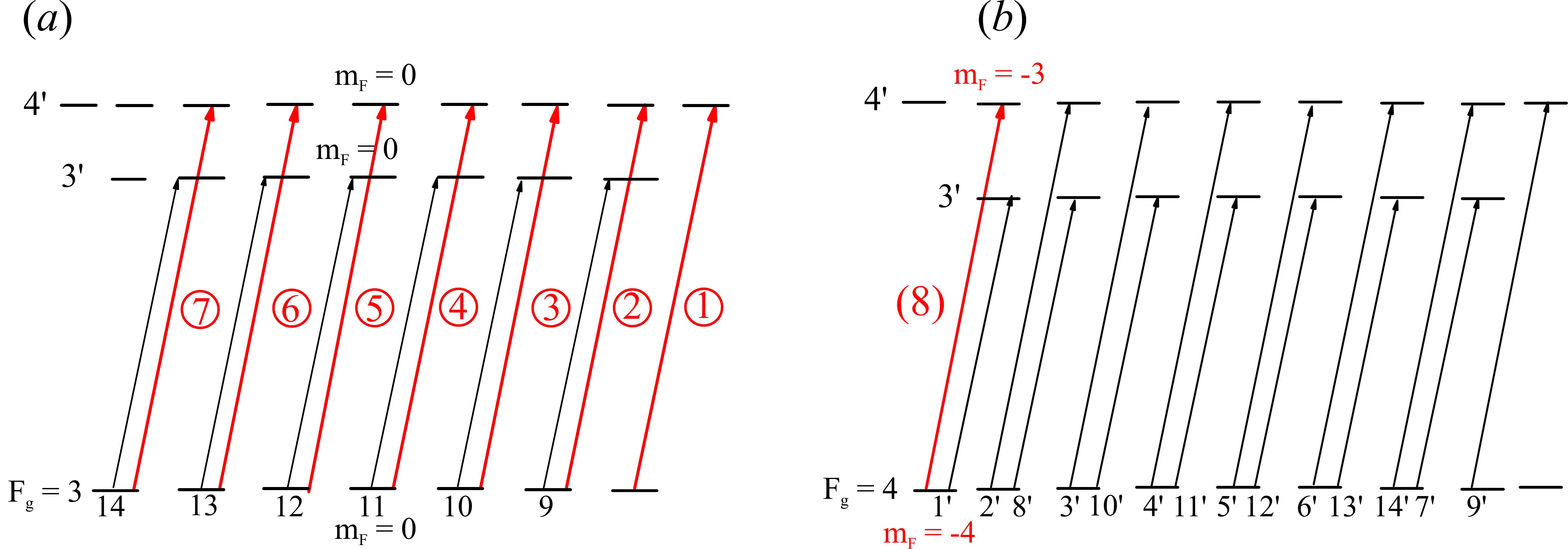}
		\caption{\label{fig:figure1} Diagram of the allowed transitions between the individual magnetic (Zeeman) sublevels for $F_g=3 \rightarrow F_e=3,4$ (\textit{a}), and $F_g=4 \rightarrow F_e=3,4$ (\textit{b}) hyperfine transitions of Cs $D_1$ line excited by $\sigma ^+$- polarized laser radiation (unprimed and primed numerals mark the lower and upper levels, respectively). Transitions labeled \textcircled{\textit{1}}--\textcircled{\textit{8}} remain in the spectra for hyperfine Paschen-Back regime (among those only \textit{8} remains from $F_g=4\rightarrow F_e=3,4$ group).}
	\end{center}
\end{figure}

All these studies benefited from the following features of nanocell: i) sub-Doppler spectral resolution for the thickness of atomic vapor nanocell $L = \lambda$ and $L = \lambda/2$ ($\lambda$ being the resonant wavelength of $D_1$ or $D_2$ line), needed to resolve a large number of individual Zeeman transition components in transmission or fluorescence spectra; ii) possibility to apply a strong magnetic field using permanent magnets: in spite of high gradient of the $B$-field reaching up to 150 G/mm, the variation of field inside atomic vapor is negligible because of the small thickness.
Among the applications based on thin atomic vapor cells placed in strong magnetic field are: i) frequency reference based on permanent magnets and micro- and nanocells, widely tunable over the range of several gigahertz by simple displacement of the magnet \cite{Sargsyan1,Sargsyan2}; ii) optical magnetometers with micro- and nanometric spatial resolution \cite{Sargsyan7,Sargsyan8}; iii) compact optical isolator using a 1 mm$^3$ heated isotopically pure $^{87}$Rb vapor cell \cite{Weller1}; iv) a simple technique to measure atomic refractive index based on Faraday rotation signal \cite{Weller2}; v) widely tunable narrow optical resonances for a diode laser radiation frequency locking \cite{Sargsyan9,Zentile1}.

In \cite{Olsen} changes in the ground-state populations of Cs $D_1$ line vapor induced by optical pumping at fixed strong magnetic field of 27 kG have been measured. However, up to date there are no sufficient data for atomic transitions of Cs $D_1$ line in wide range of applied magnetic fields. 

In the present study we have employed selective reflection (SR) technique (reflection of the laser radiation from the interface of nanocell's sapphire window and resonant Cs atomic vapor) to study behavior of transitions between magnetic sublevels of the hyperfine structure of Cs $D_1$ in external magnetic field.

The SR technique using 1 -- 10 cm-long cells is known to be an efficient  spectroscopic tool, particularly, owing to formation of a sub-Doppler spectrum in the case of low to moderate density vapors \cite{Papageorgiou,Nienhuis,Badalyan}. On the other hand, the SR allows studying the collisional processes in dense and superdense vapors \cite{Guo}. The SR technique was successfully used to study van der Waals interaction of atoms with the dielectric window of the cell, manifested by a redshift of the SR frequency (see \cite{Bloch} and references therein). In \cite{Gazazyan}, the dispersion shape of the SR signal was used to stabilize the frequency of a cw laser by locking to atomic resonance line. Several new applications of SR technique implemented for Rb nanocell $D_1$ line were recently reported in \cite{Sargsyan10}.

Below we demonstrate that the technique based on the SR from a nanocell provides several benefits as compared with earlier used techniques based on transmission or absorption spectra for the vapor column thickness $L = \lambda$ and $L = \lambda/2$. This allowed us to successfully study all the transitions between magnetic sublevels of hyperfine structure of Cs $D_1$ line in a wide range of magnetic fields (0 -- 6 kG). We present for the first time the results of theoretical and experimental studies, obtained with high spectral resolution.

\section{Theoretical model}
\label{sec:theoretical}

\begin{figure*}[h!]
	\centering
	\begin{center}
		\includegraphics[width=350pt]{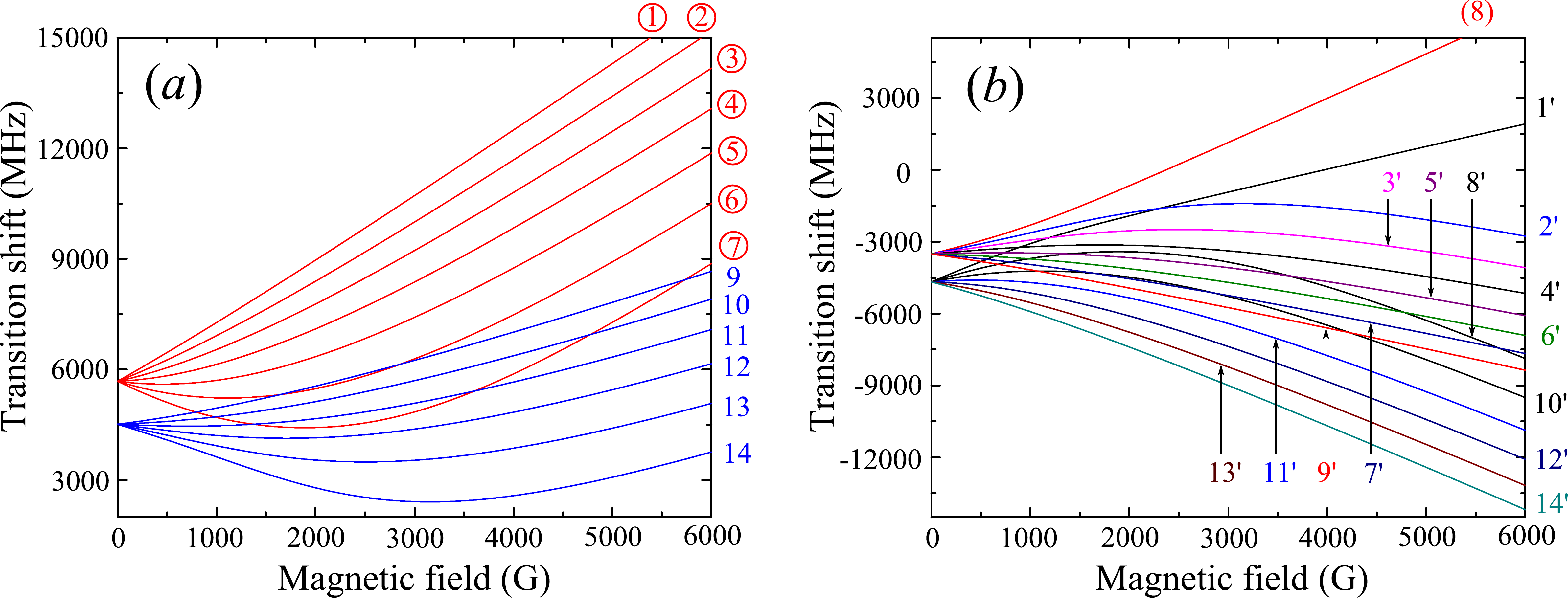}
		\caption{\label{fig:figure2} Calculated magnetic field dependence of the frequency shifts for 13 Zeeman components of $F_g=3 \rightarrow F_e=3,4$ transitions (\textit{a}) and 15  Zeeman components of $F_g=4 \rightarrow F_e=3,4$ transitions (\textit{b})  of Cs $D_1$ line in the case of $\sigma ^+$ excitation.}
	\end{center}
\end{figure*}

Our simulations of magnetic sublevels energy and relative transition probabilities for $F_g=3,4 \rightarrow F_e=3,4$ transitions of Cs $D_1$ line are based on the calculation of dependence of the eigenvalues and eigenvectors of the Hamilton matrix on a magnetic field for the full hyperfine structure manifold. This model is well known and described, particularly, in \cite{Tremblay,Aleksandrov,Auzinsh,Sargsyan3}. The allowed transitions between magnetic sublevels of hyperfine states for Cs $D_1$ line in the case of $\sigma ^+$ (left circular, $\Delta m_F = +1$) polarized excitation are depicted in Fig. \ref{fig:figure1}. The reason to consider namely the $\sigma ^+$ excitation is because in this case the positive frequency shift in a $B$-field can reach $\sim$ 20 GHz, which may be important for some applications. 

\begin{figure*}[tb]
	\centering
	\begin{center}
		\includegraphics[width=350pt]{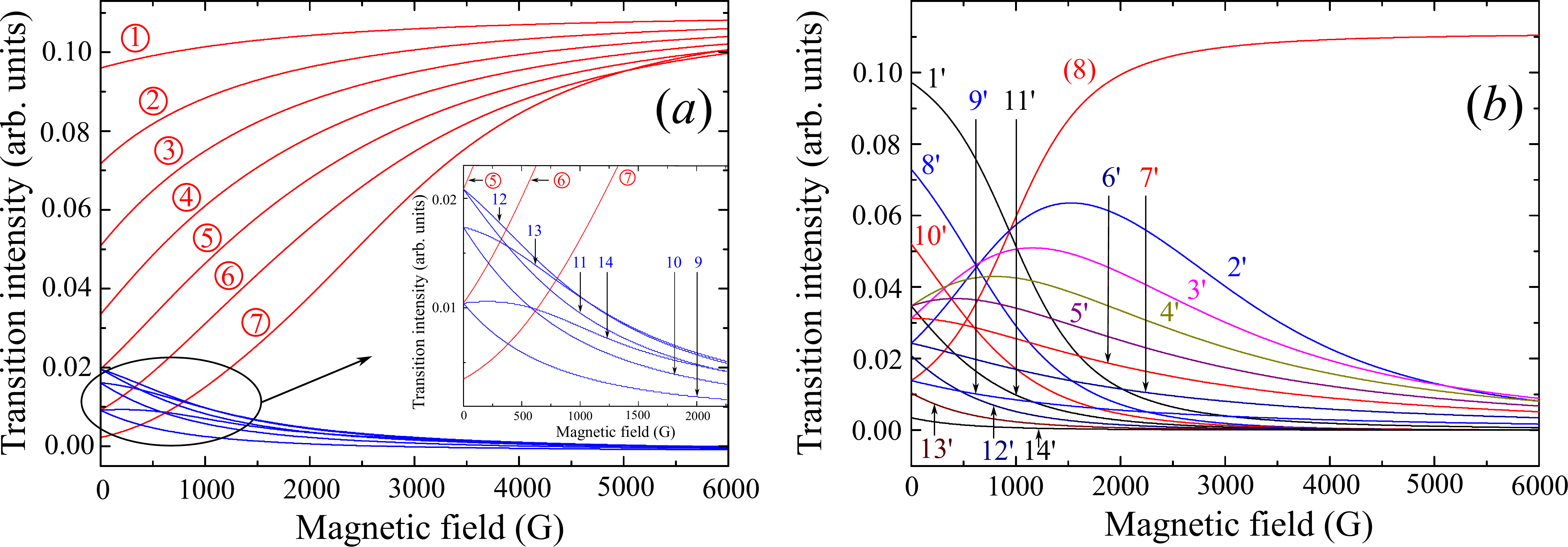}
		\caption{\label{fig:figure3} Calculated probabilities of 13 Zeeman components of $F_g=3 \rightarrow F_e=3,4$ transitions (\textit{a}) and 15  Zeeman components of $F_g=4 \rightarrow F_e=3,4$ transitions (\textit{b}) of Cs $D_1$ line versus magnetic field for the case of $\sigma ^+$- polarized excitation. Labeling corresponds to Fig. \ref{fig:figure1}.}
	\end{center}
\end{figure*}

In the case of $\sigma ^+$ excitation in relatively weak magnetic field ($\sim$ 1 kG) there are 28 allowed Zeeman transitions, among those 13 belong to $F_g=3 \rightarrow F_e=3,4$ transitions, and 15 belong to $F_g=4 \rightarrow F_e=3,4$. However, with the increase of magnetic field towards hyperfine Paschen-Back regime, only transitions labeled \textcircled{\textit{1}}--\textcircled{\textit{8}} remain in the spectra.

Calculated magnetic field dependences of the frequency shifts and probabilities of individual Zeeman components of $F_g=3,4 \rightarrow F_e=3,4$ hyperfine transitions of Cs $D_1$ in the case of $\sigma ^+$ excitation are presented in Fig. \ref{fig:figure2} and Fig. \ref{fig:figure3}, respectively. In both figures graphs (\textit{a}) and (\textit{b}) show the results for thirteen individual Zeeman components of $F_g=3\rightarrow F_e=3,4$ and fifteen individual Zeeman components of $F_g=4\rightarrow F_e=3,4$, correspondingly.

These figures show that transition \textcircled{\textit{1}}, having constant frequency slope \textit{s} = 1.8 MHz/G and high nearly invariable probability over the whole $B$-field range, is the most suitable reference for magnetic field measurements. The probabilities of transitions \textcircled{\textit{1}}--\textcircled{\textit{7}} of $F_g=3\rightarrow F_e=3,4$ group increase with $B$, while the probabilities of all other transitions (\textit{9}--\textit{14}) strongly decrease, practically down to zero for $B$ > 4 kG (see Fig. \ref{fig:figure3}\textit{a}). In the $F_g=4\rightarrow F_e=3,4$ group, the transition \textcircled{\textit{8}}, which joins \textcircled{\textit{1}}--\textcircled{\textit{7}} to form an 8-transition structure in Paschen-Back regime at high magnetic fields, exhibits distinct behavior. Being located 9.2 GHz away from \textcircled{\textit{1}}--\textcircled{\textit{7}} at $B = 0$, this transition undergoes the most rapid frequency change ($s$ = 1.85 MHz/G when approaching to \textcircled{\textit{7}} at $B > 6~$kG), and increase of probability as the $B$-field increases. The probabilities of all other transitions of this group decrease as $B$ increases over $1.5~$kG. For \textit{2'}--\textit{5'} this decrease is observed after initial growth to maximum value, and for all others the probability decreases continuously from $B = 0$. In fact, only transitions \textit{2'}--\textit{7'} remain detectable at $B > 3~$kG.

We summarize here-after the main ingredients, see reference \cite{Dutier} from Dutier \textit{et al}.,  needed to calculate the SR and absorption spectra in the case of a nanometric thickness cell. By writing the propagation equations of the electric fields inside the cell, one obtains that the reflected and transmitted signals given respectively by
%The theoretical SR and Abs. signal presented on Figs.\ref{fig:figure5}, \ref{fig:figure7}, \ref{fig:figure10} are calculated using the following formulas \cite{Dutier} 
\begin{subequations}
\small
\begin{align}
\label{eq:Sr}
S_r & \approx  2 t_{10} E_{in} \Re\Big\lbrace r\big[1-\exp(-2ikL)\big] I_{SR}\Big\rbrace/|Q|^2~, \\
\label{eq:St} 
S_t & \approx   2 t^2_{01}t_{10} E_{in} \Re\big\lbrace I_T\big\rbrace/|Q|^2 ,
\end{align}
\end{subequations}
where $t_{10}$, $t_{01}$, $r$ are  respectively transmission and reflection coefficients, $Q=1-r^2\exp(2ikL)$ is the quality factor associated to the nanocell and $I_{SR}$, $I_T$ are integrals of the induced polarization in the vapor that can be expressed, in the linear regime of interaction, as
\begin{subequations}
\small
\begin{align}
\label{eq:I_SR}
I_{SR}&=\big[1+r^2\exp(2ikL)\big]I^{lin}_{SR}-2r\exp(2ikL)I^{lin}_{T},\\
\label{eq:I_T}
I_T&=\big[1+r^2\exp(2ikL)\big]I^{lin}_{T}-2rI^{lin}_{SR},
\end{align}
\end{subequations}
with
\begin{subequations}
\small
\begin{align}
\label{eq:linear_I_SR}
I^{lin}_{SR}=C\int_{-\infty}^{+\infty}W(v)h(\omega-\omega_{eg}, \Gamma,L,v)dv, \\
\label{eq:linear_I_T}
I^{lin}_{T}=C\int_{-\infty}^{+\infty}W(v)g(\omega-\omega_{eg}, \Gamma,L,v)dv,
\end{align}
\end{subequations}
where $v$ is the speed of the atoms moving in the cell, and $W(v)$ is the assumed Maxwellian distribution of speed defined as $W(v)=(1/u\sqrt{\pi})\exp(-v^2/u^2)$ with $u$ the thermal velocity given by $u(T)=\sqrt{2k_BT/m}$ ($T$ is the temperature of the vapor, $k_B$ the Boltzmann's constant and $m$ the atomic mass). The expression of $h$ and $g$ can be found in \cite{Dutier}, they are function of the transition frequency $\omega_{eg}$, the cell thickness $L$ and of $\Gamma$ which corresponds to the optical width of the transition and include the natural  $\Gamma_{\text{nat}}$, collisional $\Gamma_{\text{col}}$ and inhomogenous $\Gamma_{\text{inhom}}$ broadening coefficients. In (\ref{eq:linear_I_SR}) and (\ref{eq:linear_I_T}), $C$ is a function of the atomic density $N$ and of the dipole moment of the transition from a state $|g\rangle$ to a state $|e\rangle$ and reads
\begin{align}
C= \frac{N t_{10}E_{in}}{4\hbar\epsilon_0Q}|\langle e|D_q|g\rangle|^2,
\end{align}
where $q$ is the standard component of the dipole associated to the scanning field polarization. The dipole moment of the transition $|g\rangle\rightarrow|e\rangle$ for an atom interacting with a longitudinal magnetic field is proportional \cite{Tremblay} to
\begin{equation}
\small |\langle e|D_q|g\rangle| \propto \sum_{F_e,F_g}c_{F_e'}c_{F_e}a(F_e,m_{F_e};F_g,m_{F_g};q)c_{F_g'}c_{F_g}~,
\end{equation}
with 
\begin{equation}
\small
\begin{split}
a(F_e,m_{F_e}&;F_g,m_{F_g};q)=(-1)^{1+I+J_e+F_e+F_g-m_{Fe}}\\
&\times\sqrt{2J_e+1}\sqrt{2F_e+1}\sqrt{2F_g+1}\\
&\times\begin{pmatrix}
F_e & 1 & F_g \\
-m_{F_e} & q & m_{F_g}
\end{pmatrix}
\left\lbrace\begin{matrix}
F_e & 1 & F_g \\
J_g & I & J_e
\end{matrix}\right\rbrace,
\end{split}
\end{equation}
where the parentheses and the curly brackets denote the 3-$j$ and 6-$j$ coefficients, respectively. 

To obtain the theoretical curves presented on Figs. \ref{fig:figure5}, \ref{fig:figure7}, \ref{fig:figure10}, we evaluated numerically Eqs. (\ref{eq:linear_I_SR}), (\ref{eq:linear_I_T}) taking into account the experimental values of $T$, $L$ and $B$. The result of the previous numerical integration was injected in Eqs. (\ref{eq:I_SR}), (\ref{eq:I_T}) and then in Eqs. (\ref{eq:Sr}), (\ref{eq:St}). Finally, these calculations are repeated over the laser frequency detuning $\omega$ in order to obtain the reflected and transmitted spectra.
\begin{figure}[h!]
	\centering
	\begin{center}
		\includegraphics[width=250pt]{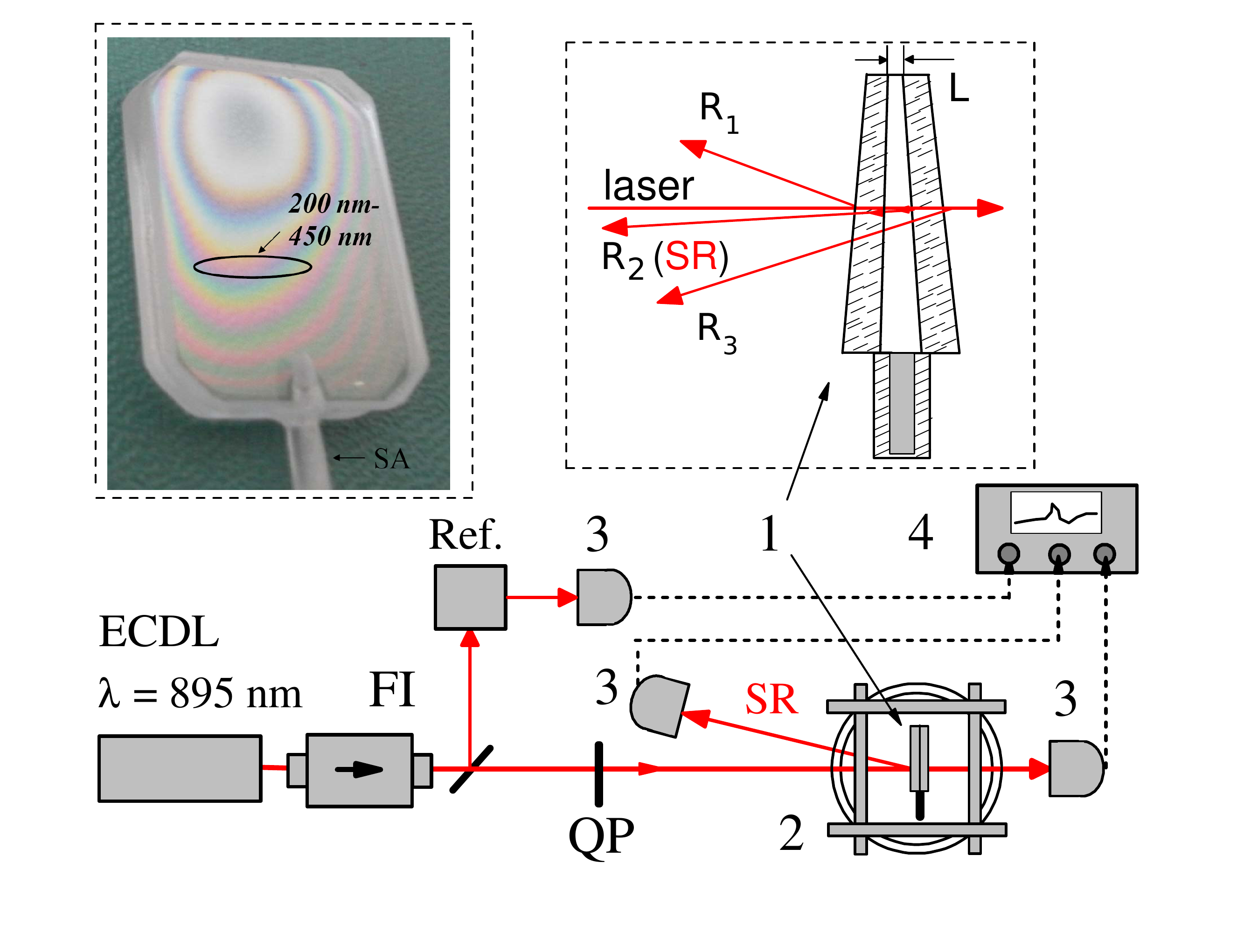}
		\caption{\label{fig:figure4} Sketch of the experimental setup. \textit{ECDL} -- extended cavity diode laser; \textit{FI} -- Faraday isolator; \textit{QP} -- $\lambda/4$ plate; \textit{1} -- Cs nanocell inside the oven; \textit{2} -- Helmholtz coils; \textit{3} -- photodetector; \textit{Ref}. -- auxiliary frequency reference unit; \textit{4} -- oscilloscope. Left inset: the photograph of nanocell; the exploited region with $L = 200 - 450$ nm is marked by an oval. Right inset: geometry of the three beams reflected from the windows of nanocell; the selective reflection (\textit{SR}) beam is labeled $R_2$.}
	\end{center}
\end{figure}
\section{Experimental}
\label{sec:experimental}

\subsection{Experimental arrangement}
\label{exparr}

\begin{figure*}[h!]
	\centering
	\begin{center}
		\includegraphics[width=350pt]{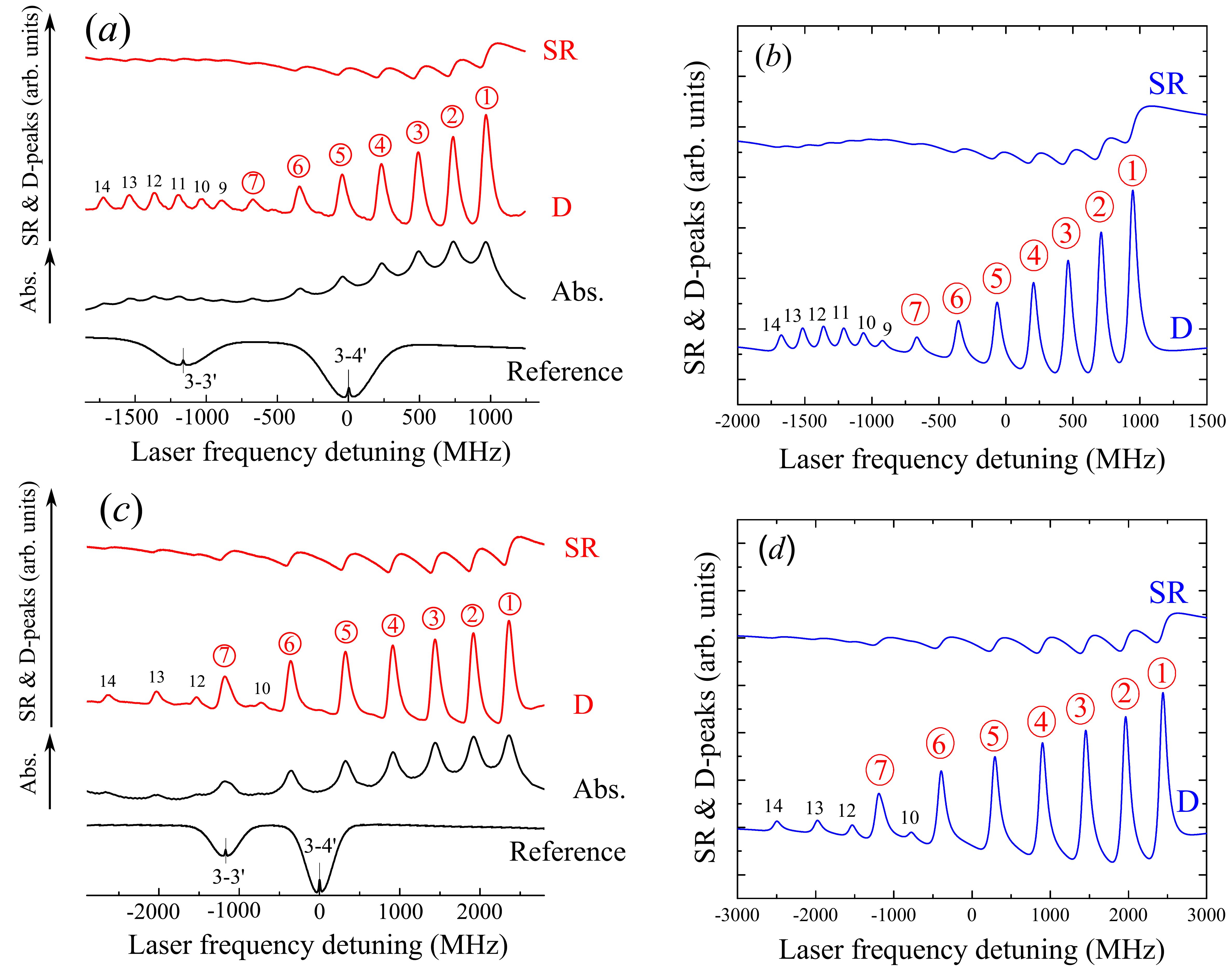}
		\caption{\label{fig:figure5} (\textit{a}), (\textit{c}) -- Experimental spectra of selective reflection (SR), its derivative (D), and absorption (Abs.) recorded under excitation of $F_g=3\rightarrow F_e=3,4$ transitions of Cs $D_1$ line by 0.1 mW $\sigma ^+$- polarized laser radiation in longitudinal magnetic field $B = 0.6$ kG (\textit{a}) and 1.5 kG (\textit{c}). The nanocell thickness $L\approx 300$ nm, the temperature 110 $^{\circ}$C. Numbering of the D-peaks corresponds to Fig. \ref{fig:figure1}\textit{a}. The bottom graph: reference spectrum. (\textit{b}), (\textit{d}) -- Calculated SR and D-peaks spectrum corresponding to experimental conditions of (\textit{a}) and (\textit{c}), respectively.}
	\end{center}
\end{figure*}

Figure \ref{fig:figure4} shows the layout of the experimental setup. To record selective reflection spectrum, the radiation of a frequency-tunable cw narrowband (1 MHz) extended cavity diode laser (\textit{ECDL}) with $\lambda = 895$ nm wavelength followed by Faraday isolator and quarter-wave plate \textit{QP} to form $\sigma ^+$ radiation was directed at normal incidence onto a Cs nanocell \textit{1} mounted inside the oven. The nanocell was placed in the center of 3-axis Helmholtz coils assembly (\textit{2}) allowing application of weak to moderate magnetic field in any direction (in present experiment $\boldsymbol{B}$ was directed longitudinally, along the laser radiation propagation direction $\boldsymbol{k}$). Stronger longitudinal magnetic field (up to $6~$kG) was applied using permanent neodymium-iron-boron alloy magnets placed near the output window of the nanocell. Variation of field strength was achieved by axial displacement of the magnet system, and was monitored by calibrated magnetometer. In spite of strong spatial gradient of a $B$-field produced by permanent magnet, the field inside the interaction region was uniform thanks to negligible thickness of the nanocell. 

In the photograph of nanocell shown in the left inset of Fig. \ref{fig:figure4} one can see interference fringes formed by light reflection from the inner surfaces of the windows because of variable thickness $L$ of vapor column across the aperture. For SR measurements (the geometry of reflected laser beams is presented in the right inset) were performed for $L = 300$ nm. The decrease of $L$ improves the spatial resolution (this is very important when using permanent magnets with high-gradient field), but simultaneously results in broadening of the SR spectral linewidth, thus 300 nm is the compromised optimal thickness.

Application of a strong magnetic field results in significant frequency shifts of detected signal (up to 10 GHz), and atomic frequency reference insensitive to applied $B$-field is needed to monitor the studied processes. To form the reference spectrum, part of the laser radiation was branched to an auxiliary saturated absorption (SA) setup with 3 cm-long Cs cell.

Selective reflection radiation, as well as nanocell transmission and SA signals were detected by FD-24K photodiodes (\textit{3}) followed by operation amplifiers. The amplified signals were simultaneously fed to Tektronix TDS2014B four-channel digital storage oscilloscope. It was recently shown in \cite{Sargsyan10} that the real time first derivative of the SR signal (hereafter referred to as D-peak) obtained while temporal scanning of laser radiation frequency yields important information on individual Zeeman transitions. D-peaks provide a low-background signal with frequency position of maximum corresponding with high precision to atomic resonance frequency, and with magnitude proportional to transition probability. This statement was done in \cite{Sargsyan10} concerning D-peaks of SR in case of Rb $D_1$ line, and has been carefully checked in the present work prior to regular measurements for the case of $F_g = 3,4\rightarrow F_e = 3,4$ hyperfine transitions of Cs $D_1$ line at zero magnetic field. In the experimental graphs below the D-peaks are presented along with directly recorded SR signals. It is worth noting that widely used saturated absorption technique \cite{Demtroder} is unsuitable for probability scaling because of its strongly non-linear nature.

All the experimental measurements presented in this article were done in the following invariable experimental conditions: the power of $\sigma ^+$- polarized laser radiation incident upon $L = 300~$nm nanocell was $P_{las} = 0.1~$mW; the nanocell temperature was $T = 110~^{\circ}$C, which is practically the same as the temperature needed to record SR from 1 -- 10 cm-long cells \cite{Papageorgiou,Nienhuis}.

\begin{figure}[h!]
	\centering
	\begin{center}
		\includegraphics[width=240pt]{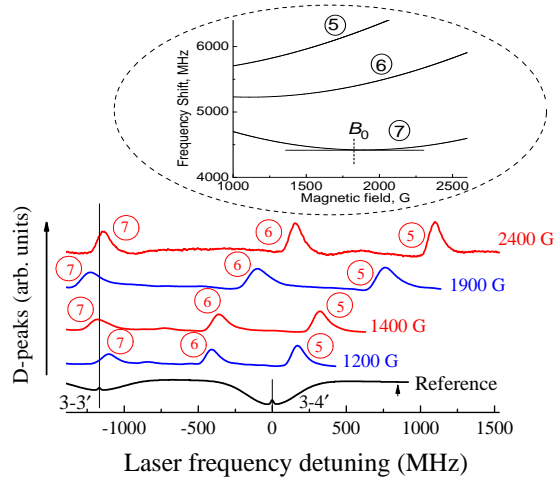}
		\caption{\label{fig:figure6} Experimental D-peaks spectra for transitions \textcircled{\textit{5}}, \textcircled{\textit{6}} and \textcircled{\textit{7}} in the case of $\sigma ^+$ radiation for $B =$ 1.2, 1.4, 1.9, and 2.4 kG. The inset shows frequency shifts of  \textcircled{\textit{5}}--\textcircled{\textit{7}} transitions as a function of magnetic field.}
	\end{center}
\end{figure}

\begin{figure*}[h!]
	\centering
	\begin{center}
		\includegraphics[width=380pt]{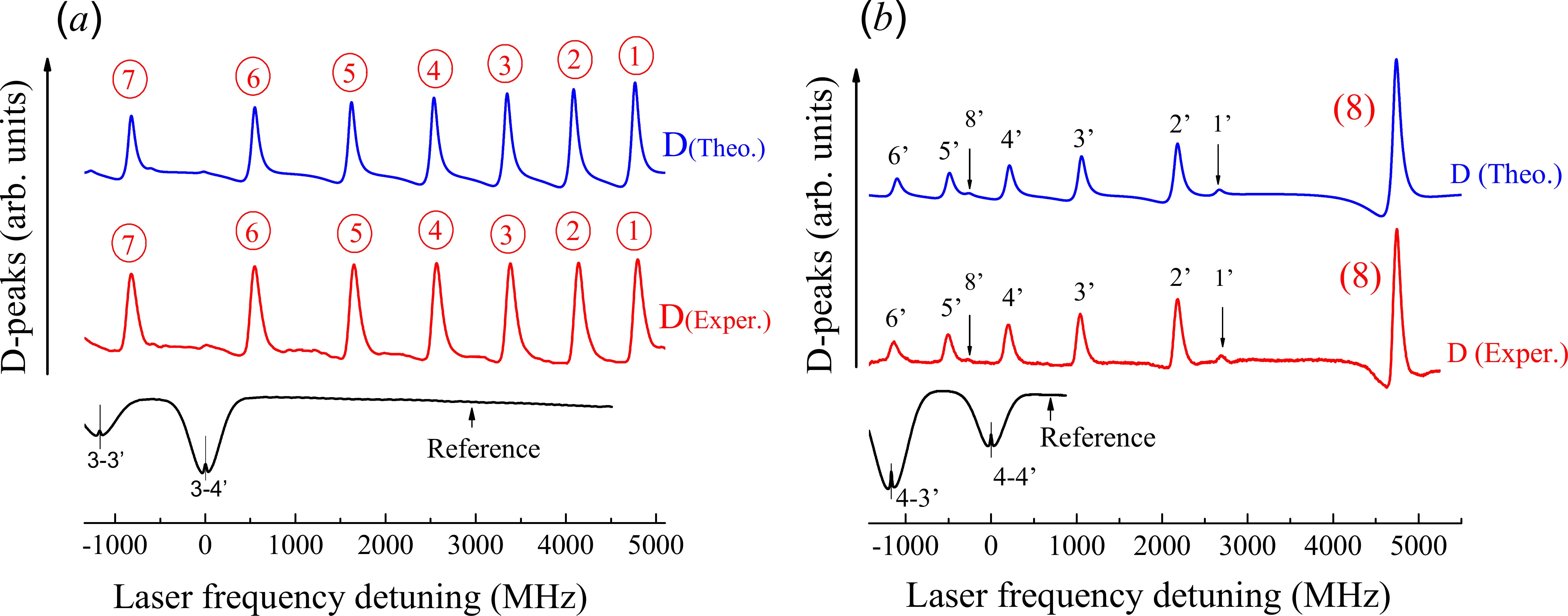}
		\caption{\label{fig:figure7} Spectra of SR derivative (D-peaks) for $F_g=3 \rightarrow F_e=3,4$ (\textit{a}) and $F_g=4 \rightarrow F_e=3,4$ (\textit{b}) transition groups of Cs $D_1$ line for a $L \approx 300$ nm nanocell excited by  $\sigma ^+$- polarized radiation in longitudinal magnetic field $B = 3$ kG. Upper trace: theory; middle trace: experiment; lower trace: reference spectrum.}
	\end{center}
\end{figure*}

\subsection{Experimental results and discussion}
\label{expresdis}

Experimentally recorded spectra for the case of longitudinal magnetic field $B = 0.6~$kG and $B = 1.5~$kG are presented in Fig. \ref{fig:figure5}\textit{a} and Fig. \ref{fig:figure5}\textit{c}, respectively. The comparison of simultaneously recorded spectra of selective reflection, its derivative (D-peaks) and absorption shows that the best contrast and the narrowest linewidth is obtained with D-peaks. As is clearly seen, all the thirteen Zeeeman transitions marked according to labeling in Fig. \ref{fig:figure1}\textit{a} are completely resolved on the D-peak trace. For normal incidence of the laser beam onto nanocell windows, the spectral line-width of D-peaks is 40 -- 50 MHz FWHM, which is nearly 10 times less than the Doppler broadening at $110~^{\circ}$C, and more than twice narrower than for the absorption spectrum ($\sim 100~ $MHz), where  transitions are resolved partially. Spectra of SR and D-peaks calculated for the parameters presented in Fig. \ref{fig:figure5}\textit{a} and Fig. \ref{fig:figure5}\textit{c} using theoretical dependences for frequency shifts and probabilities (Fig. \ref{fig:figure2}\textit{a} and Fig. \ref{fig:figure3}\textit{a}, respectively) are presented in Fig. \ref{fig:figure5}\textit{b} and Fig. \ref{fig:figure5}\textit{d} correspondingly, showing good agreement with the experiment.

As is known, the splitting of atomic hyperfine levels in weak magnetic field (as is the case for Fig. \ref{fig:figure5}) is described by a total momentum of the atom $\boldsymbol{F} = \boldsymbol{J} + \boldsymbol{I}$ and its projection $m_F$, where $\boldsymbol{J} = \boldsymbol{L} + \boldsymbol{S}$ is the total momentum of the electron, and $\boldsymbol{I}$ is the nuclear spin. The increase of magnetic field leads to the onset of hyperfine Paschen-Back (HPB) regime caused by decoupling of $\boldsymbol{J}$ and $\boldsymbol{I}$, and the splitting of levels is described by the projections $m_J$ and $m_I$ \cite{Olsen}. For $^{133}$Cs, the HPB regime occurs at $B \gg B_0 = A_{HFS}/\mu_B \approx 1700$ G, where $A_{HFS}$ is the hyperfine coupling constant for 6S$_{1/2}$, and $\mu_B$ is the Bohr magneton \cite{Zentile2}. The onset of HPB regime is manifested by: i) reduction of the quantity of Zeeman transitions to a fixed number, ii) tending the frequency slopes $s$ of grouped individual transitions to the same value; iii) equalization of individual transitions probabilities within the same group; iv) simple analytic expression for calculation of energy of the ground and excited levels (see Eq.(1) in \cite{Sargsyan2}).

\begin{figure}[tb]
	\centering
	\begin{center}
		\includegraphics[width=240pt]{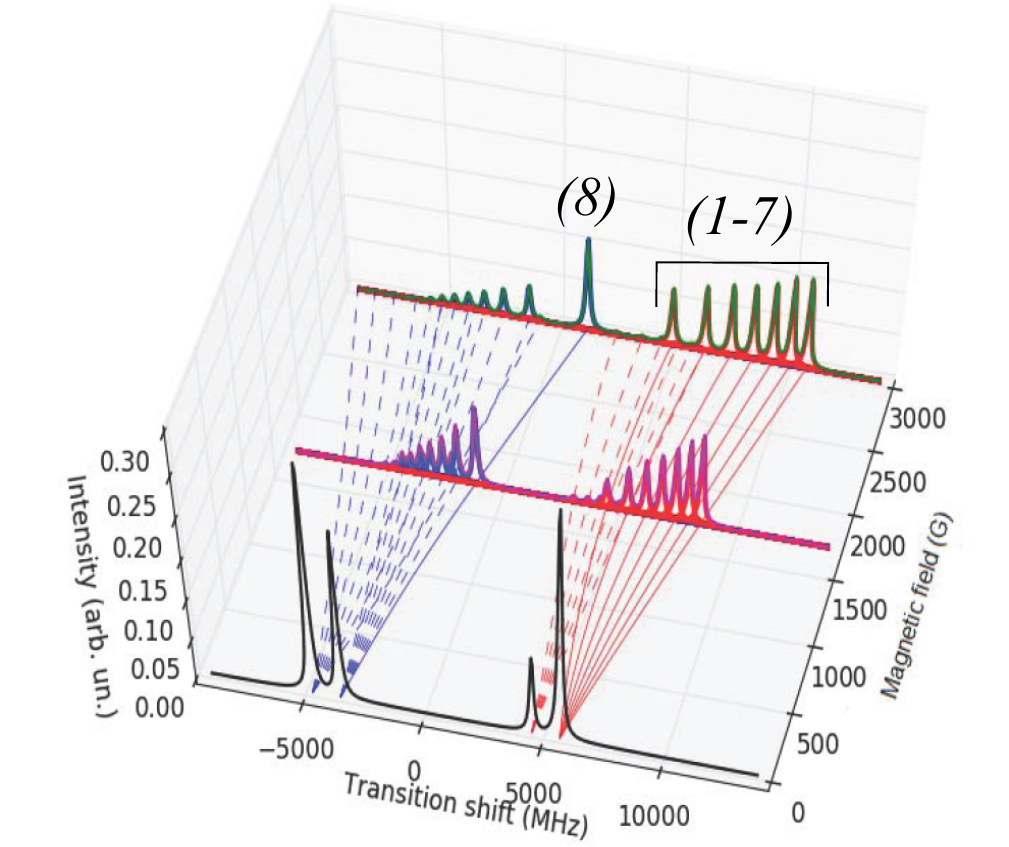}
		\caption{\label{fig:figure8} 3D presentation of calculated dependence of D-peaks spectra on magnetic field for Cs $D_1$ line. The four peaks observable at $B = 0$ correspond to $F_g=4 \rightarrow F_e=3,4$ and $F_g=3 \rightarrow F_e=3,4$ transitions (from left to right).}
	\end{center}
\end{figure}

According to theoretical model presented in Section \ref{sec:theoretical}, the frequency slopes of the  transitions labeled \textcircled{\textit{1}}--\textcircled{\textit{7}} become positive for $B > B_0$, and tend to the same value $s = 1.85$ MHz/G as for transition \textcircled{\textit{8}} with the further increase of magnetic field to $B\gg B_0$ (see Fig. \ref{fig:figure2}\textit{a}). This behavior is confirmed experimentally, as is seen in Fig. \ref{fig:figure6} presenting the frequency shifts for D-peaks labeled \textcircled{\textit{5}}, \textcircled{\textit{6}} and \textcircled{\textit{7}} for four values of magnetic field: 1.2, 1.4, 1.9 and 2.4 kG. The frequency shifts of transitions \textcircled{\textit{5}} and \textcircled{\textit{6}} continuously increase with $B$ (theoretical curves are presented in the inset), while the frequency shift of transition \textcircled{\textit{7}} increases acquiring positive slope ($s$ > 0) only at $B > B_0$, after an initial descent below the $F_g=3 \rightarrow F_e=3$ transition frequency. 

\begin{figure}[h!]
	\centering
	\begin{center}
		\includegraphics[width=200pt]{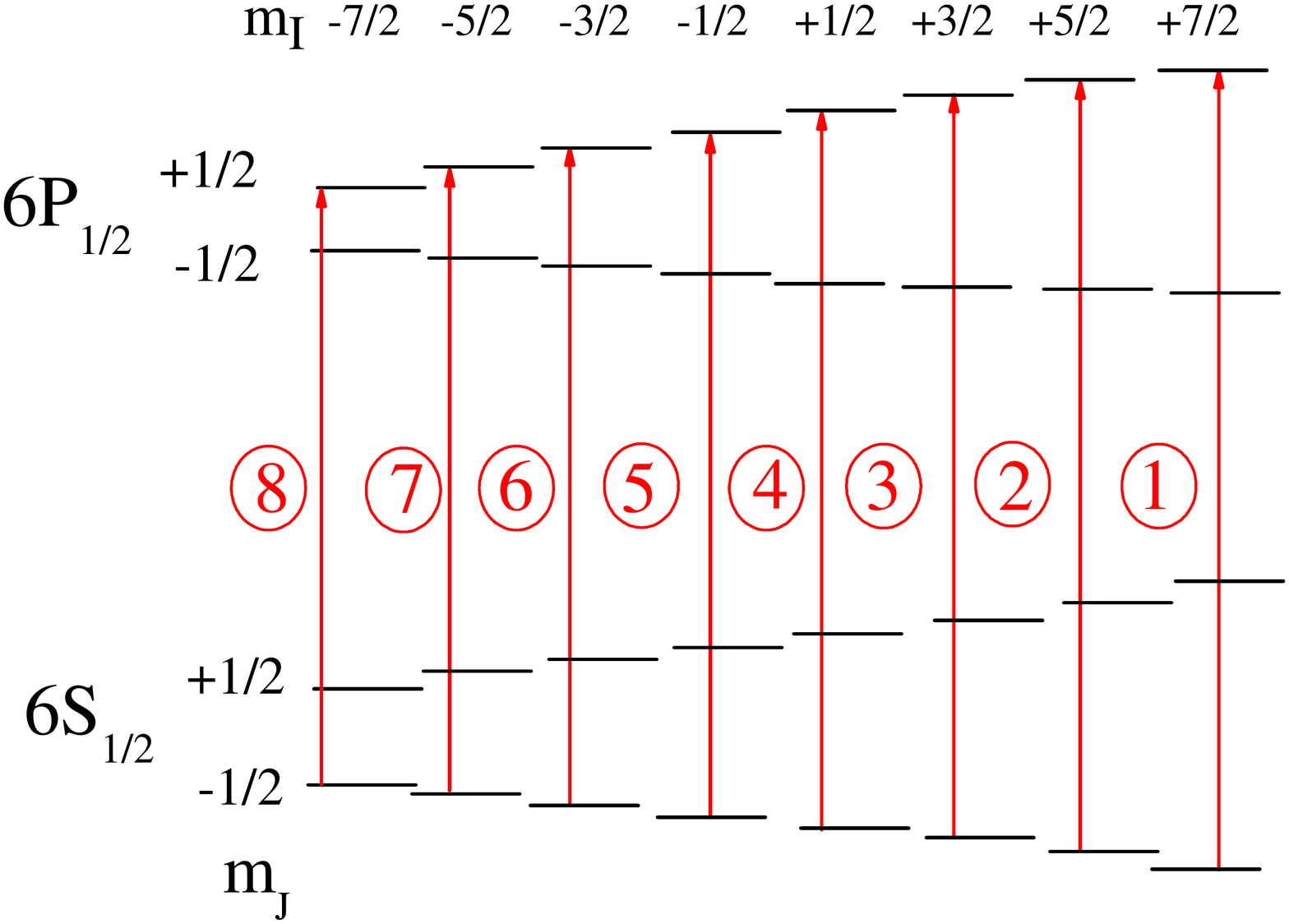}
		\caption{\label{fig:figure9} Diagram of the hyperfine structure of Cs $D_1$ line in the HPB regime. Selection rules for transitions under $\sigma^+$ excitation are $\Delta m_J =+1$; $\Delta m_I = 0$.}
	\end{center}
\end{figure}

Theoretical and experimental spectra of D-peaks for a longitudinal magnetic field $B \approx 3$ kG are shown in Fig. \ref{fig:figure7}. At this intermediate magnetic field the transition \textcircled{\textit{8}} is still too far from the "HPB group" \textcircled{\textit{1}}--\textcircled{\textit{7}}, and it was not possible to cover all these transitions in a single spectrum (the smooth mode-hop-free scanning range of our ECDL is 5 -- 6 GHz). For this reason the spectra are shown separately for the regions of $F_g=3 \rightarrow F_e=3,4$ (\textit{a}) and $F_g=4 \rightarrow F_e=3,4$ (\textit{b}) transitions of Cs $D_1$ line (the lower traces in the graphs are the reference SA spectra). Good agreement between the recorded and calculated spectra indicates that the modeling (see Figs. \ref{fig:figure2}, \ref{fig:figure3}) properly describes also the case of intermediate $B$-field, which is the most complicated for treatment. It should be noted that even weak transitions labeled \textit{1'} and \textit{8'} are well resolved. 

A 3D view showing the evolution of calculated D-peak spectra with the increase of magnetic field is presented in Fig. \ref{fig:figure8}. The plots are built using theoretical curves for frequency shifts and probabilities presented in Figs. \ref{fig:figure2}, \ref{fig:figure3}. As is seen, Zeeman transitions \textcircled{\textit{1}}--\textcircled{\textit{7}} forming the HPB transition structure at $B \gg B_0$, emerge from $F_g=3\rightarrow F_e=4$ transition, while the last Zeeman transition of the HPB octet, \textcircled{\textit{8}} emerges from $F_g=4\rightarrow F_e=4$. The probabilities of all other transitions undergo gradual suppression with the increase of $B$.

As was mentioned above, the onset of HPB regime is characterized by breaking of coupling between $\boldsymbol{J}$ and $\boldsymbol{I}$, and the splitting of atomic levels is described by projections $m_J$ and $m_I$. The diagram of transitions in HPB regime of Cs $D_1$ line allowed by selection rules for $\sigma ^+$ excitation is shown in Fig. \ref{fig:figure9}.

\begin{figure}[h!]
	\centering
	%\begin{center}
		\includegraphics[width=260pt]{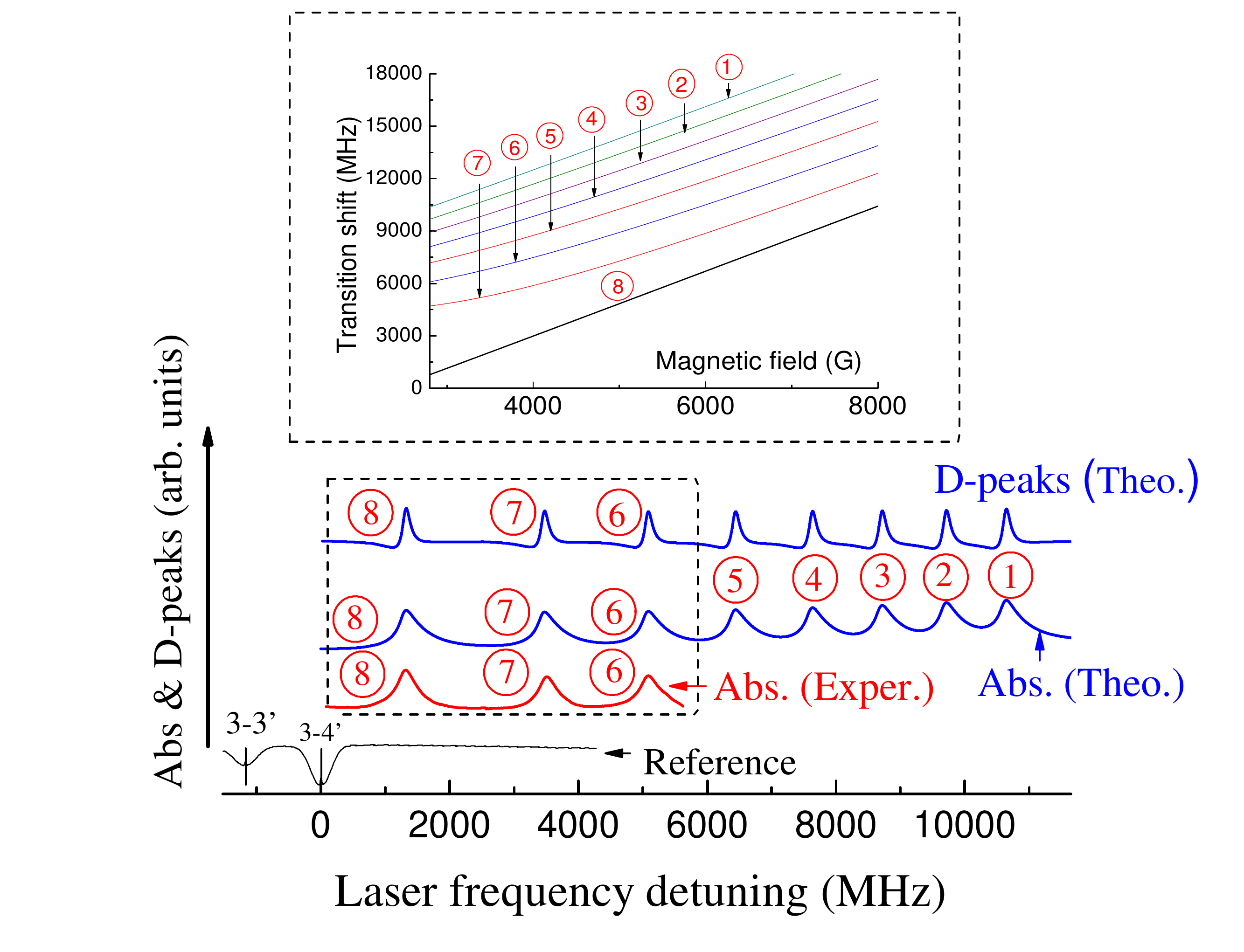}
		\caption{\label{fig:figure10} Spectra for $\sigma ^+$ excitation of Cs $D_1$ line in longitudinal magnetic field $B = 6~$kG. Upper trace: calculated D-peaks spectrum in the region of \textcircled{\textit{1}}--\textcircled{\textit{8}} transitions; middle trace: calculated absorption spectrum (blue) and fragment of the experimentally recorded absorption spectrum (red) covering the region of \textcircled{\textit{6}}--\textcircled{\textit{8}} transitions; lower trace: reference SA spectrum. The inset shows frequency shifts of \textcircled{\textit{1}}--\textcircled{\textit{8}} transitions as a function of magnetic field. }
	%\end{center}
\end{figure}

Calculated D-peaks spectrum showing \textcircled{\textit{1}}--\textcircled{\textit{8}} transitions for $B$ = 6 kG, the maximum attainable value in our experiment, is presented in Fig. \ref{fig:figure10}, along with experimentally recorded fragment of absorption spectrum covering the region of \textcircled{\textit{6}}--\textcircled{\textit{8}} components. As it is seen from the calculated frequency shift dependence depicted in the inset of Fig. \ref{fig:figure10}, transition \textcircled{\textit{8}} joins the group \textcircled{\textit{1}}--\textcircled{\textit{7}} at $B\ge 6~$kG, which is confirmed by experimental spectrum.

In order to create such high magnetic filed by simple means, we used a system presented in Fig. 2 of \cite{Sargsyan2}. The assembly consists of two strong permanent ring magnets mounted on a $50\times50~$mm cross-section $\Pi$-shaped holder made from soft stainless steel. The system with small inlet and outlet radiation holes was designed for transmission of a small diameter laser beam, and it was not suitable for selective reflection configuration. For that reason we presented in Fig. \ref{fig:figure10} absorption but not SR D-peaks experimental spectrum. In the figure the upper trace shows calculated D-peaks spectrum for  \textcircled{\textit{1}}--\textcircled{\textit{8}} transitions, meanwhile the middle trace shows the calculated absorption spectrum. Though the spectral linewidth of absorption lines is broader than the linewidth of D-peaks (see Fig. \ref{fig:figure5}), the absorption lines in the experimental graph correctly display the frequency position and amplitudes (probabilities) of \textcircled{\textit{6}}--\textcircled{\textit{8}} transitions. In particular, the modeling predicts that the amplitude of the transition labeled \textcircled{\textit{8}} must be 1.1 times larger than for \textcircled{\textit{7}}, which is consistent with the experimental observation. 

From the experimental graphs presented in Figs. \ref{fig:figure5} -- \ref{fig:figure8}, \ref{fig:figure10} demonstrating the evolution of spectra with the increase of magnetic field from $0.6~$kG to $6~$kG, it is clearly seen that the transitions probabilities of \textcircled{\textit{1}}--\textcircled{\textit{8}} components tend to the same value for $B \gg B_0 \approx 1.7~$kG, while all other transitions gradually reduce to zero, manifesting establishment of hyperfine Paschen-Back regime.

In \cite{Dutier} it was shown that the nanocell behaves as a low-finesse Fabry-P{\'e}rot etalon, which determines the ratio of intensities of the reflected radiations $R_2/R_1$ (see Fig. \ref{fig:figure4}). Since for the case of normal incidence of laser radiation employed in our experiment the beams $R_2$ (SR) and $R_1$ propagate in nearly the same (backward) direction, the following procedure was used to extract the net SR signal (some percents of incident radiation power, depending on nanocell temperature). After each measurement, the laser radiation frequency was tuned to off-resonance area, and the recorded signal was subtracted from the resonant one. 

Preliminary results show that it is still possible to detect the SR and D-peaks at the nanocell thickness less than 100 nm with some worsening of frequency resolution and contrast (it results in broadening of D-peak up to 150 MHz and reduces the reflection caused by frequent collisions of the Cs atoms with the nanocell's windows). Using selective reflection from nanocell with $L < 100~$nm, it could be possible to provide mapping of strongly inhomogeneous magnetic fields, such as very high gradient fields (40 kG/mm) used in \cite{Folman}. We have checked also that the shape of the SR derivative D-peak in the case of 1-cm long cell filled with the Cs is not suitable for precise magnetic field measurements.

\section{Conclusions}

We have presented for the first time the experimental and theoretical studies of 28 Zeeman transitions of Cs $D_1$ exposed to wide range of external magnetic field (0 -- 6 kG). High spectral resolution has been achieved by implementing selective reflection of $\sigma ^+$- polarized laser radiation from a cesium vapor nanocell with $L \sim 300~$nm thickness of vapor column. It is shown that real time derivative of SR signal (D-peak), having 40 MHz (sub-Doppler) linewidth, located at the atomic transition frequency and having amplitude proportional to transition probability, can be successfully used as a spectroscopic tool to monitor the transformation of Zeeman splitting to hyperfine Paschen-Back splitting with the increase of the applied magnetic field. We have shown that from 28 individual Zeeman transitions observed at low magnetic fields only 8 remain in the spectrum when $B = 6~$kG, manifesting establishment of hyperfine Paschen-Back regime. The obtained experimental results are in a good agreement with theoretical modeling throughout the whole $B$-field range.

Thanks to relatively intensity of the SR signal (several percents from the incident radiation), its linear response, and low divergence of the reflected beam, SR from a nanocell can be used for high-distance remote monitoring and mapping of both homogeneous and highly inhomogeneous magnetic fields in a wide range with sub-micrometric spatial resolution. Besides, the proposed technique can be also used as a precise frequency reference for atomic transitions. We should note that recent development of a glass nanocell operating at lower temperature ($\sim 100~^{\circ}$C) \cite{Whittaker} can make the SR technique available for wider range of researchers and developers.

\section*{Funding Information}

This work was supported by the RA MES State Committee of Science, in the frames of the research projects No. 15T-1C040 and No. 15T-1C277.

We are grateful to A.S. Sarkisyan for fabrication of the nanocells.

The research was conducted in the scope of the International Associated Laboratory IRMAS (CNRS-France \& SCS-Armenia).

\end{document}